\documentclass{article}
\usepackage{lineno}
\usepackage[utf8]{inputenc}
\usepackage{textgreek}
\usepackage{geometry}
\usepackage{longtable}
\usepackage{array}
\usepackage{booktabs}
\usepackage{authblk}
\usepackage{xcolor}
\usepackage{subcaption}
\usepackage{graphicx}
\usepackage{listings}
\usepackage[colorlinks=true, linkcolor=blue, citecolor=blue, urlcolor=blue]{hyperref}
\title{Leveraging NCBI Genomic Metadata for Epidemiological Insights: Example of  \textit{Enterobacterales}}

\author{ Bryan Harris,  Majid Bani-Yaghoub\footnote{Corresponding author. Email: baniyaghoubm@umkc.edu}}
    \affil{\small  Division of Computing, Analytics \& Mathematics, School of Science and Engineering, University of Missouri-Kansas City, 5100 Rockhill Rd., Kansas City, Missouri 64110, USA}
  \date{}

\begin{document}
\maketitle

\begin{abstract}

Numerous studies have utilized NCBI data for genomic analysis, gene annotation, and identifying disease-associated variants, yet NCBI’s epidemiological potential remains underexplored. This study demonstrates how NCBI datasets can be systematically leveraged to extract and interpret infectious disease patterns across spatial and temporal dimensions.
Using \textit{Enterobacterales} as a case study, we analyzed over 477,000 genomic records and metadata, including collection date, location, host species, and isolation source. We compared trends of \textit{Escherichia coli} and \textit{Salmonella} in NCBI data with CDC’s National Outbreak Reporting System (NORS). While both datasets showed consistent seasonal peaks and foodborne sources, NCBI data revealed broader host species (e.g., wildlife, environmental reservoirs), greater isolate diversity, and finer spatial-temporal resolution.
These insights were enabled by our open-source Python package, EpiNCBI\_V1, developed for real-time downloading, filtering, and cleaning of pathogen genomic metadata from NCBI. This work highlights the value of integrating genomic repositories into public health analytics to enhance surveillance, outbreak detection, and cross-species transmission tracking globally.

\end{abstract}

\section{Introduction}
  The National Outbreak Reporting System (NORS) is a system governed by the Center for Disease Control (CDC) as a modern-day means for tracking and reporting of enteric disease outbreaks. The NORS Dashboard was launched in 2009 as a publicly accessible platform for standardized outbreak reporting by public health departments \cite{outbreak_reporting}. NORS collects data on outbreaks driven by transmissions classified as foodborne, waterborne, person-to-person  or via contact with animals. Indeterminate/Unknown tranmission vector data are also reported to NORS. Reports are submitted on a voluntary basis by local, state, and territorial health agencies, as well as CDC officials. \cite{CDC2025NORSData}  \\

Researchers have used NORS data extensively to understand the epidemiology of different infectious diseases. 
This includes characterizing the relative contributions of different transmission modes (e.g., person-to-person, foodborne) 
and etiologies such as norovirus in enteric outbreaks (person-to-person: 62\%, foodborne: 24\%, norovirus: 59\%) 
\cite{lively2022nors}, 
estimating transmissibility through reproduction number modeling in norovirus outbreaks 
(e.g., median $R_{0}$ of 2.75, with variation by setting and season) \cite{allen2020nors}, and to estimate attribution of various pathogens in outbreaks sourced by various leafy greens as vectors of foodborne illness (e.g., an estimated 19.8\% of all \textit{Eschericha Coli} O157:H7 illnesses are attributed to Romaine lettuce.) \cite{leafygreens}.\\

In addition to NORS data, the National Center for Biotechnology Information (NCBI) offers a wealth of biological datasets.
NCBI is a unified collection of multiple genomic databases such as GenBank and RefSeq which provide reference nucleotide and protein sequences. \cite{benson2013genbank,oleary2016refseq}.
NCBI was formally established by legislation on 4 November 1998 as a division of the National Library of Medicine (NLM). Unlike NORS, any individual or institution can submit data to NCBI through NCBI's submission portal. Submissions are processed or rejected in accordance with NLM's GenBank and SRA Data Processing document \cite{NLM_data_processing}.  It offers scientists access to a large and wide variety of publicly available biological data, including genomic sequencing
data for many different organisms, including gene and genomic data, as well as viruses. \cite{ncbi}. \\
    Similar to NORS, there are many applications of NCBI data in epidemiology. For example, \textit{Campylobacter jejuni} genomic data sourced from the NCBI Pathogen Detection Database is used to understand the distribution of virulence genes in a set of \textit{Campylobacter jejuni} genomes \cite{campy_genes}. NCBI data can also be cross-referenced with other datasources such as the Food Safety and Inspection Service (FSIS) outbreak database to find similarities or differences in distributions of serovar and isolation source for \textit{Salmonella} strains \cite{Comparison_2025}. \\

In light of these examples, NORS and NCBI appear to naturally complement each other for epidemiology and infectious disease research. While some articles exist that compare and contrast NCBI data to other outbreak sources \cite{GenomeGraphR_2019}, to the best of our knowledge, no research article has  contrasted NORS and NCBI for epidemiological studies. Part of the issue is the lack of tools that can streamline the process of comparing data from these two sources. The present study aims to fill this gap. In particular, NORS data can be downloaded directly from NORS's website; however, there is a degree of difficulty involved in obtaining NCBI data in a clean and ready-to-use format.
Three main categories of data packages are currently available: genome, gene, and virus. Users can customize the contents of any data package. Genomic data can be obtained by direct download and some curated programs. Direct downloads from NCBI's website are in the form of zip files.  This enables researchers to download a full dataset of submissions for any choice of pathogen. On the other hand, the  NCBI curated programs download the data from the command-line, which allows for experienced programmers to create sophisticated programs to efficiently download, format, and filter data tailored to a specific pathogen(s) for their research needs.
For custom analytics solutions, neither of these options includes a protocol for cleaning the data. \\

While data submitted to NCBI may be corrected in the instance of clerical errors of a typographical nature, several errors of type have been observed. Although packages such as ShinyR could potentially be used for different outbreak database comparisons \cite{GenomeGraphR_2019}, there seem to be no Python packages for the purposes of comparing NCBI submission data with NORS data. With the understanding that NCBI and NORS data fit well together, we develop and test a robust automated tool in Python to assist researchers in analyzing patterns in NCBI genomic data submissions. 

We call the first version of this package, EpiNCBI, which sequentially downloads, formats, cleans, and visualizes detailed NCBI metadata such as isolation source and host, submitter, year and month of submission, and serotype. To validate the  EpiNCBI-V1 package, we apply it to NCBI data associated with \textit{E. Coli} and \textit{Salmonella} and compare the resulting analyses with corresponding NORS data. Hence, the present  study demonstrates the capabilities of the developed Python package to uncover trends, seasonality, and patterns present in NCBI data that may be overlooked in NORS data.\\

The rest of this paper is organized as follows. In Section 2, we provide the methodology used for data collection, Python tool development, and data visualization. Section 3 will illustrate the patterns that are present in the curated data sets obtained from the NCBI and NORS, the databases, when they are compared simultaneously. Section 4 rewinds the discussion of main findings and plans for further development of EpiNCBI Python package.

\section{Materials and Methods}

\subsection{Overview}

\begin{figure}[htbp]
\includegraphics[width=\linewidth]{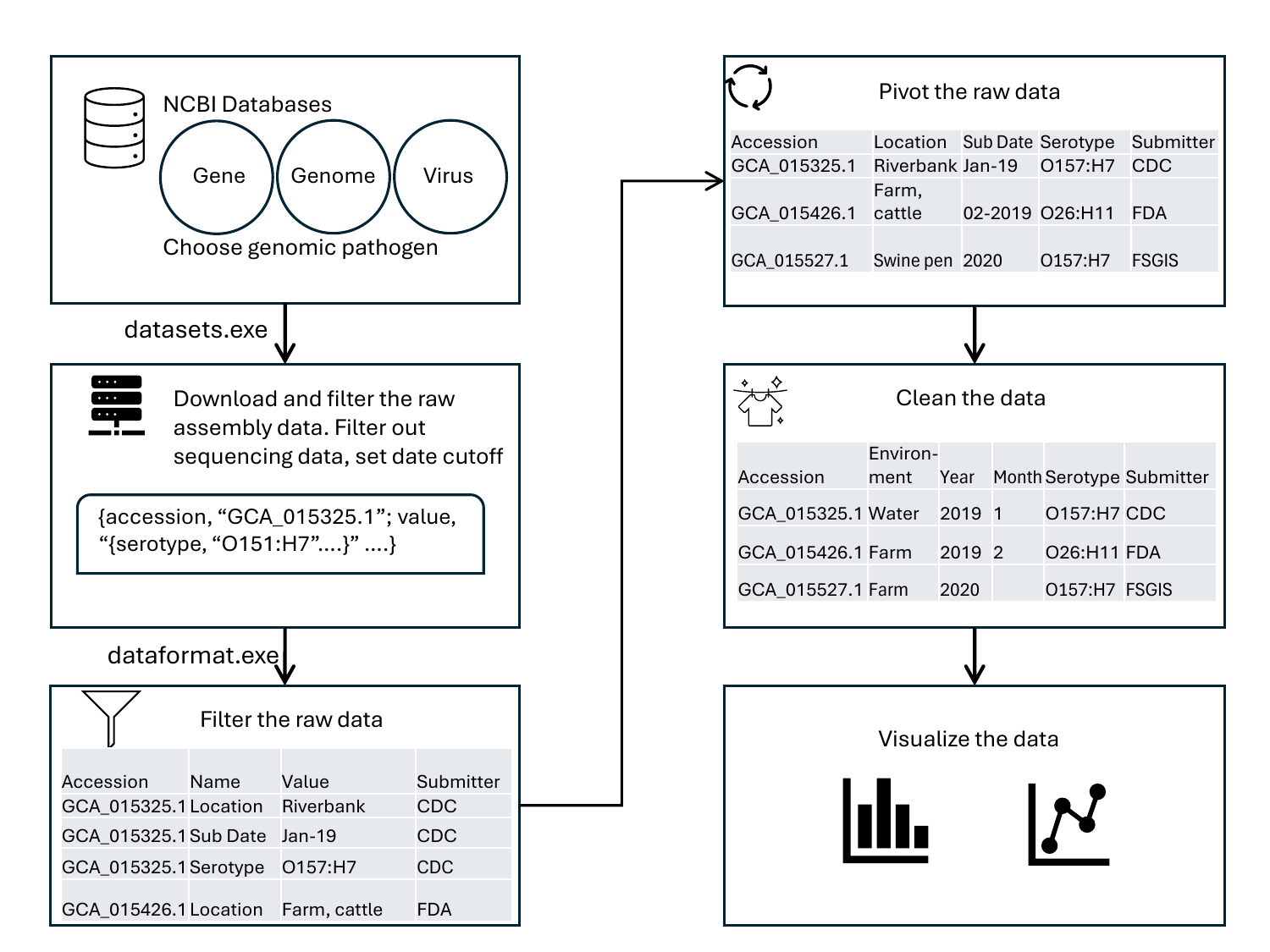}
        \caption{Pipeline for obtaining NCBI genomic   data using the developed Python tool ``EpiNCBI\_V1.}
    \label{fig1}
\end{figure}

We prepare a data pipeline to illustrate the process involved in this study. The process for obtaining and pre-processing NCBI genomic submission data is illustrated in a six-compartment diagram in Figure \ref{fig1}. The pipeline is executed by an author-designed Python package in collaboration with two tools developed by NCBI. The order of the steps is the process is choosing the data, downloading the data with filtering, formatting the data with a 2nd filtering, pivoting the data, cleaning the data, and visualizing the data alongside NORS outbreak data. We gave the name EpiNCBI\_V1 to this Python package to exemplify its use for obtaining NCBI data for epidemiological studies. The "V1" tag shows this is the first assembled version of the code, and the authors plan to expand its functionality in future work.

\subsection{Python Tool Development}
The EpiNCBI\_V1 Python Tool is developed as follows. The data analysis pipeline begins with the selection of a genomic pathogen by the user, as shown in compartment one of Figure \ref{fig1}. NCBI databases feature gene, genomic, and virus data \cite{ncbi}. For our study, we will focus only on genomic data, but it can be extended to the other two databases. To begin, EpiNCBI\_V1 requests a single NCBI-defined taxonomy ID from the user to designate the genomic pathogen they'd like data for. For example, the data used in the development of this study used \textit{Escherichia Coli} and \textit{Salmonella} taxonomy IDs 562 and 590 respectively. Other taxonomy IDs may be found directly by searching for a specific pathogen's name on the NCBI website. \\

After the user inputs a valid taxonomy ID, EpiNCBI\_V1 calls an executable  program named "Datasets", developed by NCBI. Datasets facilitates the initial download and filtering of the NCBI genomic data based on the Taxonomy ID entered by the user and date range specified. Other filters are available when calling Datasets, including options for the download of genomic sequences along with their assembly metadata. We opted to download only assembly metadata, and filtered out any sequence data that the metadata is describing. We also filter by submission date, to include only samples submitted to NCBI before 2024 to align our analysis with NORS, which only includes outbreak data up to the end of 2023. The data is then downloaded from NCBI's servers as a .jsonl file. \cite{ncbi}. \\

Because .jsonl files contain nested .json lines, they are not compatible with many data analysis programs. To remedy this, EpiNCBI\_V1 calls another NCBI-developed program named ``Dataformat". Dataformat is a tool designed to convert the .jsonl format data to either a .tsv file, or an .xlsx file for analysis in data spreadsheet programs \cite{ncbi}. Dataformat was also designed with column filters in mind. The downloaded raw data contains many more columns than were used for this study. EpiNCBI\_V1 ultilizes Dataformat to only keep four columns; accession, assminfo-biosample-attribute-name, assminfo-biosample-attribute-value, and submitter, and saves the result as a .tsv file to avoid data truncation that occurs when saved as an .xlsx file with too many rows. The 3rd panel in Figure \ref{fig1}  illustrates a sample of \textit{E. Coli} data after the Dataformat tool is applied in our study. \\

To further optimize analysis, we want to have individual rows in our .tsv file correspond to exactly one assembly submission, and have all the downloaded metadata for individual assemblies in exactly one row. We did not identify a particular tool curated by NCBI to do this task. EpiNCBI\_V1 fills this gap by pivoting the filtered raw data with the pivot\_table function, so that every field in the attribute name column becomes a new column in the formatted data, with values equal to the adjacent attribute value column entry. This leads to the creation of approximately 580 new columns of data, one for each unique entry found in the assminfo-biosample-attribute-name column,  of which we only keep eight: Geographic Location, Date of Collection, Strain, Host, Serotype, Isolation Source, and Source Type. A small sample of the data after pivoting is shown in the fourth compartment in Figure  \ref{fig1} . \\

The last step EpiNCBI\_V1 carries out in Figure \ref{fig1} is cleaning the pivoted data. We observed that the columns representing geographic location, collection date, and isolation source contain entries that offer insights in multiple areas. For time series analysis, we use a string search function to separate the collection data column into month and year. For location analysis, we use word lookup. For example, an entry in geographic location may read "river in jackson mississippi" of which we can extract the state name "Mississippi" and an environment of contamination "Water Source". Some geographic location entries only contained major cities, or state abbreviations. To infer and extract state information, we utilized a lookup table obtained from the GitHub page of Plotly, a popular module with the opensource Python package MatPlotLib commonly used for generating customized data visualization solutions. This lookup table contains 300 of the most populated cities in the US, accompanied by their state location. If a geographic location entry contains a city on this list, EpiNCBI\_V1 replaces the entry with the state name. Other lookup tables used for analysis by EpiNCBI\_V1 can be found on the author's GitHub page \cite{GitHub}. \\

\subsection{New Data Acquisition System}

With Figure \ref{fig1} established, EpiNCBI\_V1 provides a solution to accessing formatted, clean, and ready-to-use genomic assembly data in the NCBI databases with minimal effort. The source code for EpiNCBI\_V1 is assembled into a Python file, EpiNCBI\_V1.py, available at the author's GitHub page \cite{GitHub}, with additional documentation found in the ReadMe.txt file in the root of the project folder. The ReadMe file contains details for installing Python, running EpiNCBI\_V1.py locally, and a brief overview of the major functions inside EpiNCBI\_V1.py that carry out the actions of the Figure \ref{fig1}.

\subsection{NCBI-NORS Comparative Analysis}

EPiNCBI\_V1 concludes with saving the data in .csv format. The .csv files are then utilized by other software packages to perform data visualization or data-driven modeling. A few examples of data visualization are heat maps showing disease incidence across regions, time-series plots of infection dynamics, network diagrams of host–pathogen interactions, and bar/box plots comparing intervention scenarios. Examples of data-driven modeling based on the NCBI and NORS datasets are predictive models of outbreak dynamics using machine learning classifiers, regression models identifying key environmental or demographic risk factors, time-series forecasting of incidence trends, and compartmental or agent-based simulations calibrated with reported outbreak data \cite{Review2024, Babanejaddehaki2024, Gao2020,  Gao2025, Steele2019}. \\

In the present work, we compare NORS and NCBI datasets for the pathogens \textit{E. Coli} and \textit{Salmonella} using a combination of time-series, barcharts, and heatmaps to uncover new insights, while noting that EpiNCBI\_V1 is compatible with other pathogens as well not used in this study. The primary reason for choosing these two pathogens is that they are members of an important gram-negative family of pathogens responsible for frequent outbreaks and food poisoning in the US, with several cases of hospitalizations and development of antimicrobial resistance.

\section{Results}

\subsection{Overview}

The EpiNCBI\_V1 package was built using E. Coli and Salmonella as references, but is able to be applied other other organisms in the NCBI genome databases as well, for example, EpiNCBI\_1 has been verified to follow the data pipeline in  Figure \ref{fig1} for \textit{Campylobacter jejuni} (Taxonomy ID = 197) and for \textit{Listeria} (Taxonomy ID = 1637), presenting further opportunities for novel epidemiological research. With a pipeline in place for end-to-end data acquisition, we now present results for NORS and NCBI data with EpiNCBI\_V1 applied to E. Coli and Salmonella (i.e., NCBI taxonomy ID 562 and 590, respectively). Below are the spatial-temporal analyses associated with these two pathogens. \\

For \textit{E. Coli} NCBI data, the download step resulted in 281,487 submissions. Two more filters were applied submissions for both pathogens were further filtered to only include submissions collected with a geographic location containing the string "USA", "UNITED STATES", or a state name within the US, or a city name within the top 300 most populous cities in the US. This filtering showed 113,858 samples of \textit{E. Coli} within the US, and 363,207 samples of \textit{Salmonella} within the US. For a state-by-state analysis, US-based samples were further filtered based on if their geographic location contained either an exact state name or abbreviation, or contained a city that falls on the list of the top 300 most populous cities. We observed 45,913 samples of \textit{E. Coli} with state-level information, and  similarly for \textit{Salmonella} we obtained 531,912 records.

\subsection{Time Series Analysis}

The processed data can be analyzed to identify temporal trends and possible seasonality. Figure S1 in the supplementary document provides descriptive statistics of the main submitter of data associated with these two pathogens to NCBI. It can be seen that a significant portion of data is submitted by organizations other than the CDC. Furthermore, Figures S2-S5 in the supplementary document provide additional details on NORS data. This includes the number of outbreaks by top ten pathogens (Figure S2), temporal changes in the number of outbreaks reported by NORS (Figure S3), Word cloud of etiologies (Figure S4), and a comparison of the temporal pattern of outbreaks by Norovirus as the pathogen causing the highest number of outbreaks and Enterobacterales and other pathogens (Figure S5).\\

\begin{figure}[htbp]
\includegraphics[width=\linewidth]{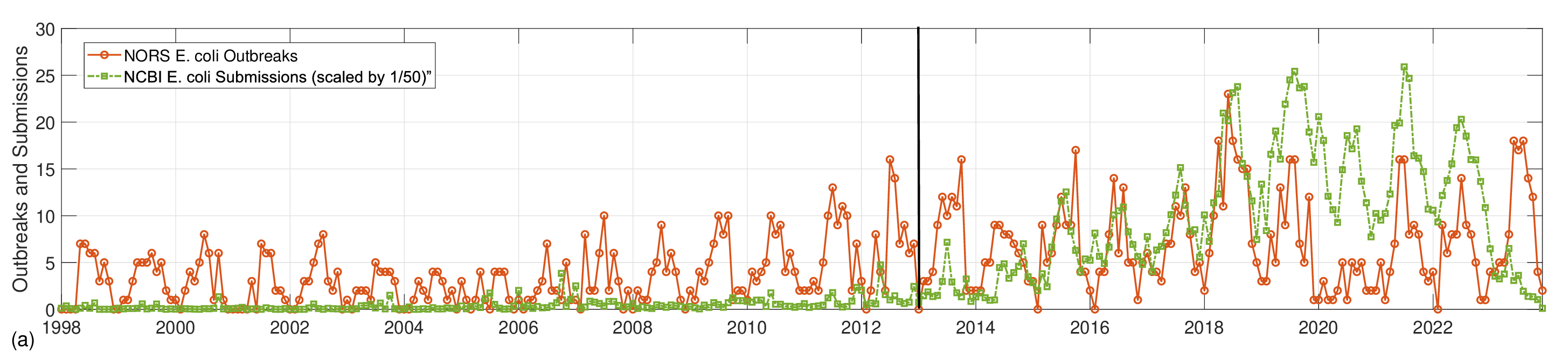}
        \includegraphics[width=\linewidth]{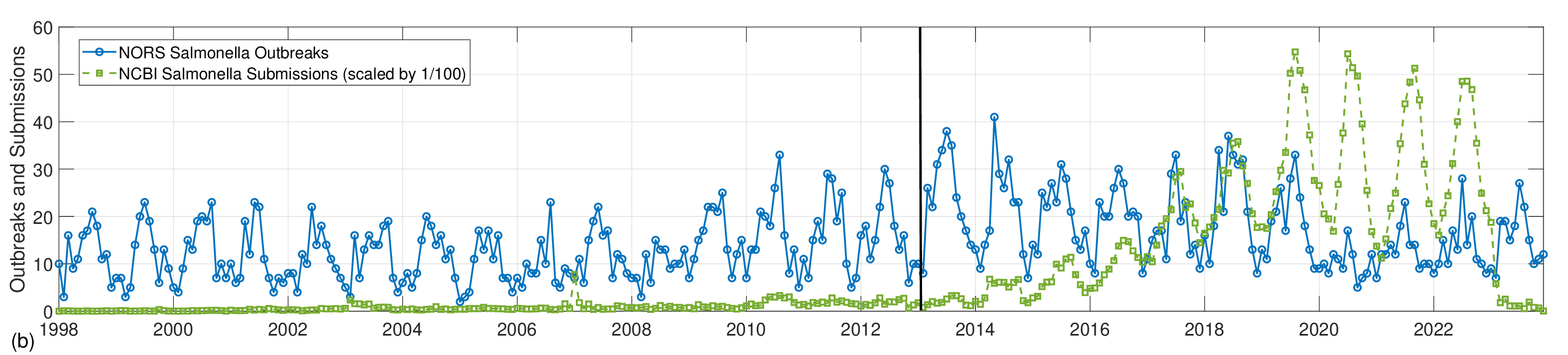}
        \caption{Time series comparisons of NCBI submission and NORS outbreaks data. Bacterial strains: (a) \textit{E.~coli} and (b) \textit{Salmonella.}}
    \label{fig:two-panels1}
\end{figure}

Figure \ref{fig:two-panels1} illustrates NORS and NCBI submission data from January of 1998 through the end of December 2023. Analysis of the time-series data shows an uptrend from 2013 to 2019, followed by a sharp increase in NCBI sample submissions during the COVID-19 pandemic (2019–2022). However, a downtrend in submissions is observed in 2023. Only NCBI samples that included a collection year and month in the metadata are counted. The NORS data follows a seasonal trend, in Summer outbreaks are more prevalent than Winter outbreaks. Before 2013, NCBI submission data counts of E. Coli and Salmonella initially are very small prior to 2013. In 2013 the number of NCBI submissions increases exponentially. This increase in submissions coincides with the launch of GenomeTrakr \cite{timme2018utilizing}. NCBI submission data for E. Coli and Salmonella also displays seasonal trends with Summer peaks and Winter lows.\\

\begin{figure}[htbp]
\includegraphics[width=\linewidth]{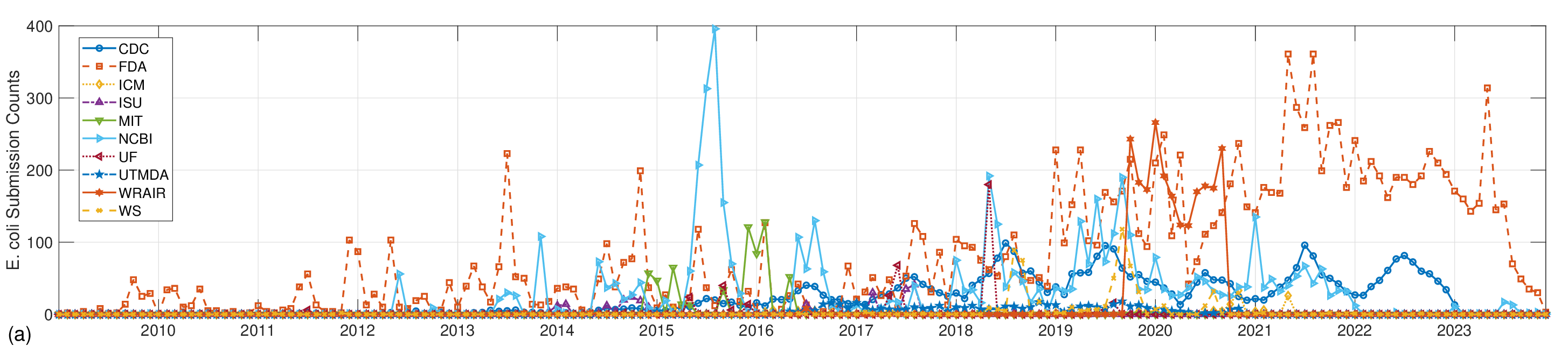}
        \includegraphics[width=\linewidth]{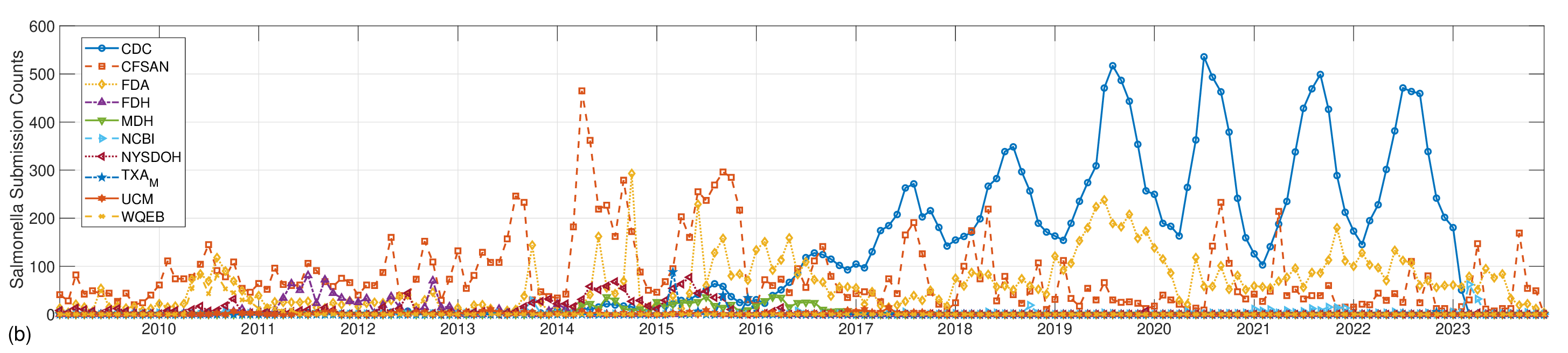}
        \caption{Time series of NCBI submission counts by top ten submitters. (CDC scaled by 1/100) (2009-2023)}
    \label{fig:two-panels}
\end{figure}

Figure S6 in the supplementary document displays cumulative bar charts for NORS and NCBI submission data at the monthly level for E. Coli and Salmonella. NCBI submissions require a collection month for inclusion in Figure S6. Namely,  72,144 NCBI records for \textit{E. Coli} and 253,304 NCBI records for \textit{Salmonella} contained submission months. Figure S6 asserts the claims that both NORS and NCBI data follow a seasonal trend, with higher numbers of reports during the Summer months, with a peak in July and a low in February. The data for both \textit{E. Coli} and \textit{Salmonella} appear to exhibit a one or two-month delay in reporting, as the maximum and minimum values for NCBI and NORS are shifted one or two months ahead of the maximum and minimum of the US average temperature curve.  Figure \ref{fig:two-panels} clearly shows that the NCBI submissions associated with \textit{Salmonella} and \textit{E. coli} have increased over the time. There are also clear seasonal patterns in the data. Note that CDC submissions have been scaled by $1/100$. In this case, some submission patterns become visible. For \textit{E. Coli} and \textit{Salmonella}, the CDC has been the top distributor until early 2023. While it is unlikely the CDC ceased submitting samples in early 2023, it is more likely that CDC either submits samples in batches, or delays reporting until the new year. Other institutions only submitted samples during certain years. Most institutions also show evidence of seasonality in their submission timelines. \\

Additional time series plots can be found in the supplementary document. Specifically, Figure S7 shows that the number of outbreaks caused by \textit{Salmonella} is two to three times higher than those caused by \textit{E. coli}.

\begin{figure}[htbp]
\includegraphics[width=\linewidth]{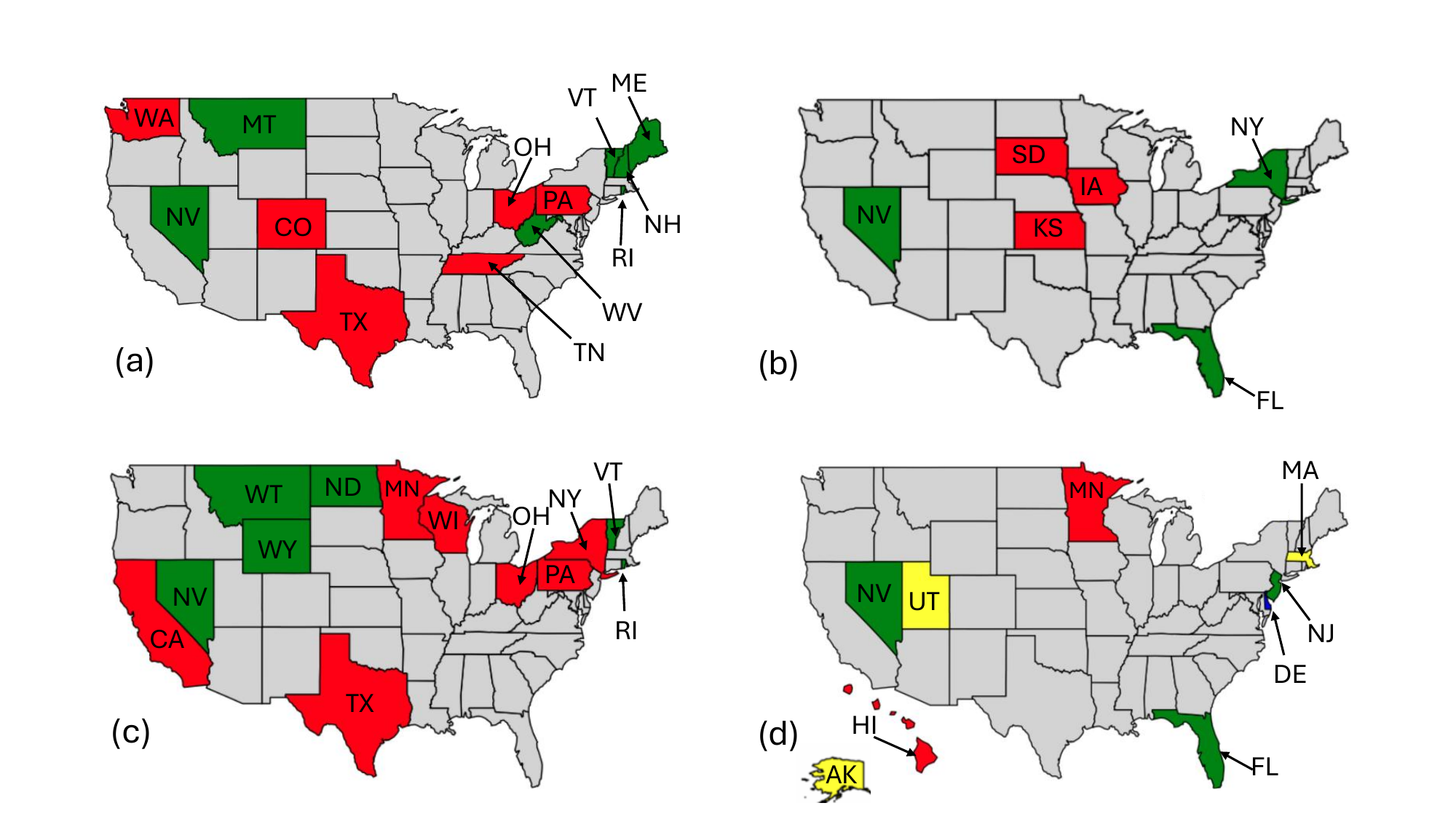}
        \caption{Overlaps between NCBI and NORS data for  \textit{E. coli} (panel a) and \textit{Salmonella} (panels c). Panels (b) and (d) correspond to data normalized by the population of each state (i.e., per capita in million). Red: overlaps between top 10 NCBI submissions and outbreak, 
        Green=overlaps of bottom 10 submissions and outbreaks, Yellow: top 11 outbreaks, bottom 10 submissions; Blue: bottom 10 outbreaks, top 10 submissions.    }
    \label{fig:maps}
\end{figure}

\subsection{Spatial Data Analysis}

The downloaded NCBI data also contains fields for isolation source and geographic location at the state level.  Figure \ref{fig:maps} depicts four maps of the US, each exhibiting areas of interest related to \textit{E. Coli} and \textit{Salmonella}. Figures \ref{fig:maps} (a) exhibits states with the highest and lowest total number of \textit{E. Coli} counts in NCBI submissions and NORS outbreak sources. Red states that are top ten in NCBI submission geographic location and top ten in NORS outbreak state of origin. Green states show states that are bottom ten in NCBI submission geographic location and bottom ten in NORS outbreak state of origin. Yellow states depict states that are bottom ten in NCBI submission geographic location and top eleven in NORS outbreak state of origin. Lastly, blue states are in the top ten in NCBI submission geographic location and in the bottom ten in NORS outbreak state of origin. Figure \ref{fig:maps} (b) shows a similar situation to Figure a), except the total submitted count for NCBI and NORS data of each state is divided by their respective population in million, exhibiting \textit{E. Coli} prevalence patterns. For example, Texas was a top 10 submitter for NCBI submissions and NORS outbreak sources, but when the number of counts and number of outbreaks are divided by the state's population, Texas is no longer one of the top ten states for submissions or outbreaks. Figures \ref{fig:maps} (c) and (d) represent the number of submissions associated with \textit{Salmonella}. \\

We can also use NCBI data to define broad categories for Chi-square testing. For this, we classified each US submission according to its isolation source and host columns, and created a new category field classifying each submission as having been derived from beef, fruit, humans, pork, turkey, or water. We conducted a Chi-square independence test to assess independence in the category-state pairings. Such information is not present in the NORS data, which can help public health officials and researchers better interpret outbreak dynamics and inform decision-making.

\begin{figure}[ht]
\begin{minipage}{0.48\textwidth}
  \centering
   \includegraphics[width=\linewidth]{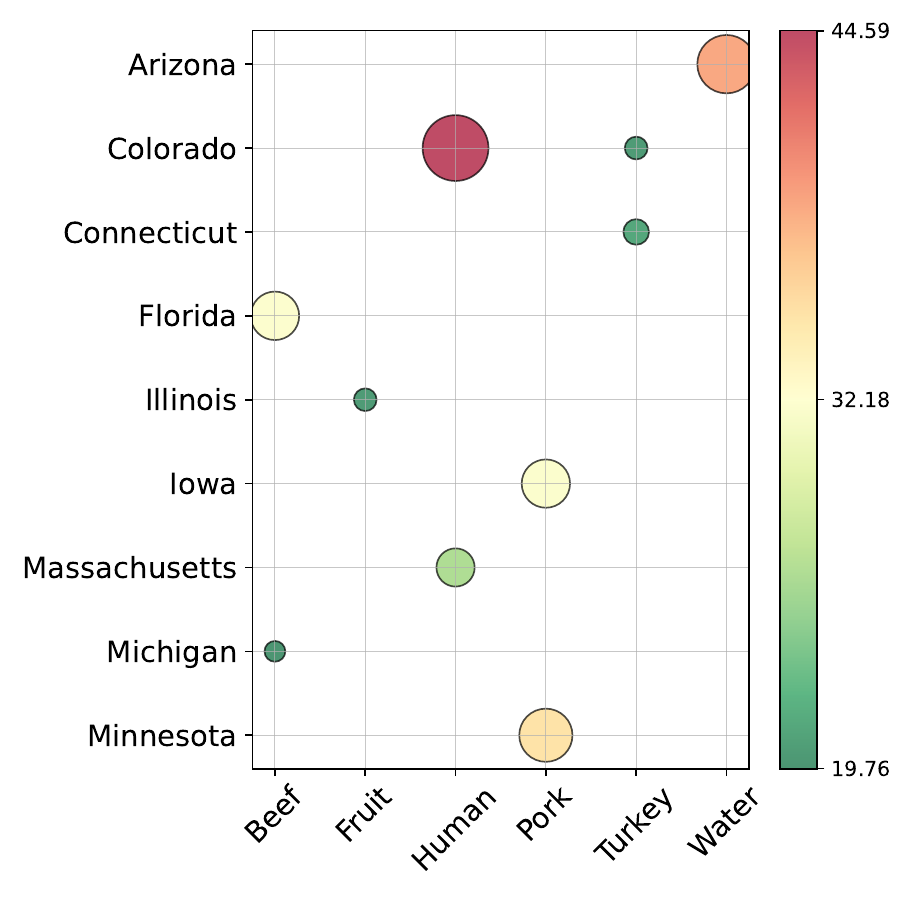}
\end{minipage}
\hfill
\begin{minipage}{0.48\textwidth}
  \centering
  \includegraphics[width=\linewidth]{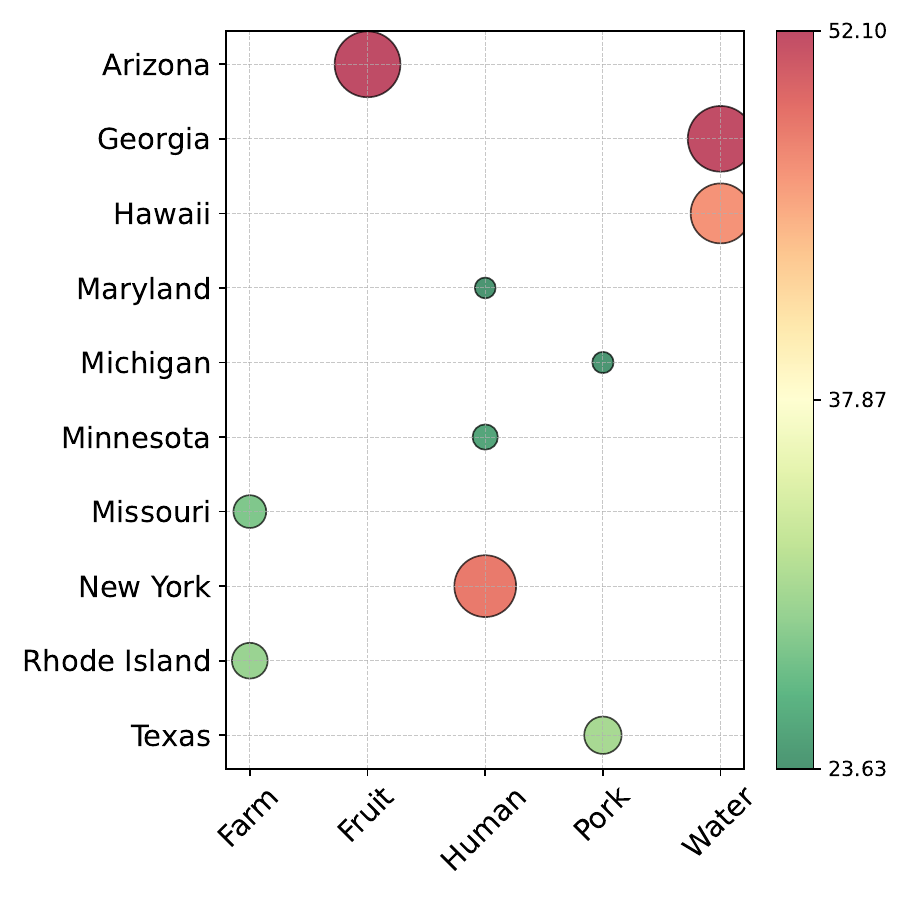}
\end{minipage}
\caption{Bubble charts for the top ten contributors to the chi-square test for independence statistic by state and source for  \textit{E. Coli} (left) and \textit{Salmonella} (right). The largest contributors are red. Smaller contributors are in green.}
\label{Fig5}
\end{figure}

\section{Discussion and Conclusion}

In the present work, we developed a Python tool (EpiNCBI V1 p) that enables researchers to access curated and standardized NCBI genomic metadata for epidemiological modeling and analyses. EpiNCBI V1 p supports a wide range of applications, including data-driven modeling, visualization of temporal and spatial trends, and evidence-based decision-making. To illustrate its utility, we compared curated NCBI data with NORS reports of \textit{Salmonella} and \textit{E. coli} outbreaks, highlighting the added value of integrating genomic and epidemiological data streams. \\

To reproduce the cleaning and analysis in this paper, the authors have prepared a GitHub page containing the programs and code scripts used to generate each dataset used to obtain the figures in this paper, as well as the supplementary document \cite{GitHub}. The code contains comments throughout highlighting the Python functions used to carry out each compartment in \ref{fig1}. Looking ahead, future developments will focus on expanding compatibility with additional surveillance platforms, enhancing real-time data processing, and strengthening applications in genomic surveillance, One Health initiatives, and rapid outbreak response.\\

A number of issues were encountered while working with NCBI datasets Sample submissions to NCBI do not have a strict format to adhere to, and many fields are voluntary. Submission dates do not always follow the same formatting, reported strains are not always consistent (e.g., O157 vs O157:H7 vs 157), and a number of other fields we are interested in require a large amount of time to be spent in cleaning the data. No two cleaning methods will yield the same results. The provided data in the fields for isolation source and host are often not specific, and require some inference to determine what source each individual sample was extracted from. (i.e., an isolation source of "urine" does not specify if the host was animal or human). Additionally, categorizing the isolation source and host columns analysis was determined using a limited number of keyword searches. \\

In conclusion, the variability and inconsistency of NCBI sample submissions highlight the challenges of relying on non-standardized metadata. Despite these limitations, careful curation and cleaning can still yield valuable insights for epidemiological analysis.

\section{Acknowledgments}

The Authors would like to acknowledge Sudhiksha Kumar (Division of Computing, Analytics \& Mathematics, School of Science and Engineering, University of Missouri-Kansas City) and Supraja Kanagala (Division of Computing, Analytics \& Mathematics, School of Science and Engineering, University of Missouri-Kansas City) for their contributions in making this work possible.

\bibliographystyle{plain}
\bibliography{ref} 

\end{document}


\maketitle
\listoffigures
\tableofcontents

\section{Descriptive Statistics of NORS and NCBI Data}
This section shows count figures for NORS and NCBI that are neither spatial nor temporal in nature.

\begin{figure}[ht]
\begin{minipage}{0.48\textwidth}
  \centering
   \includegraphics[width=\linewidth]{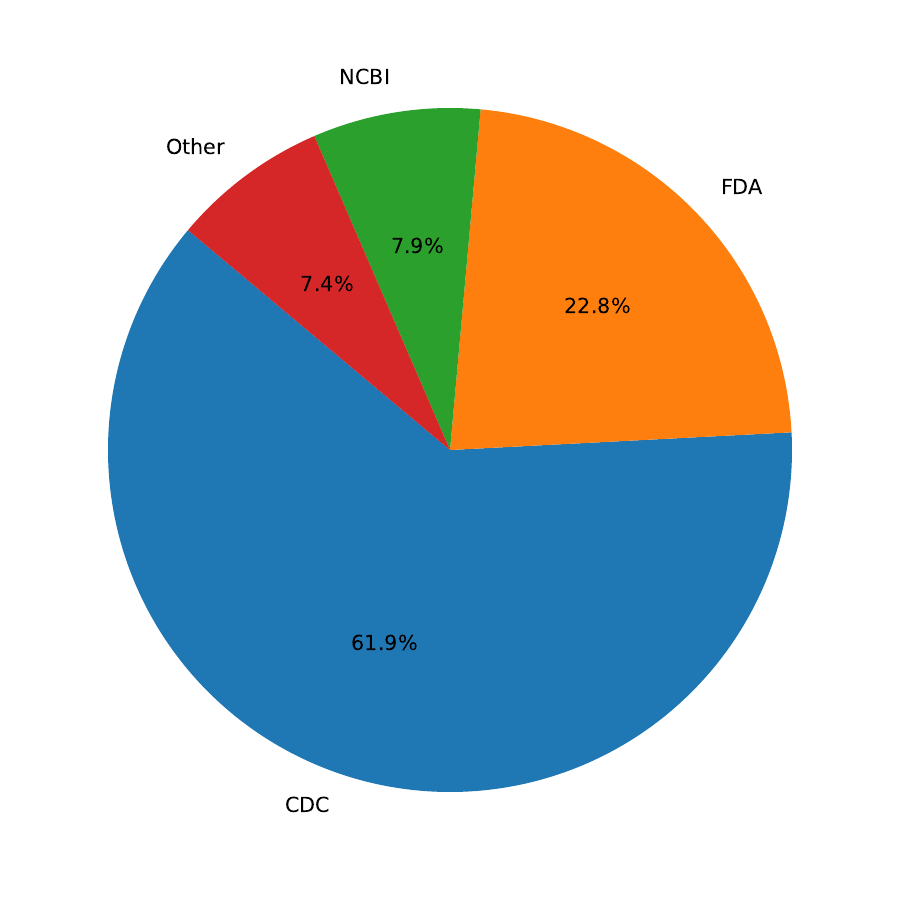}
\end{minipage}
\hfill
\begin{minipage}{0.48\textwidth}
  \centering
  \includegraphics[width=\linewidth]{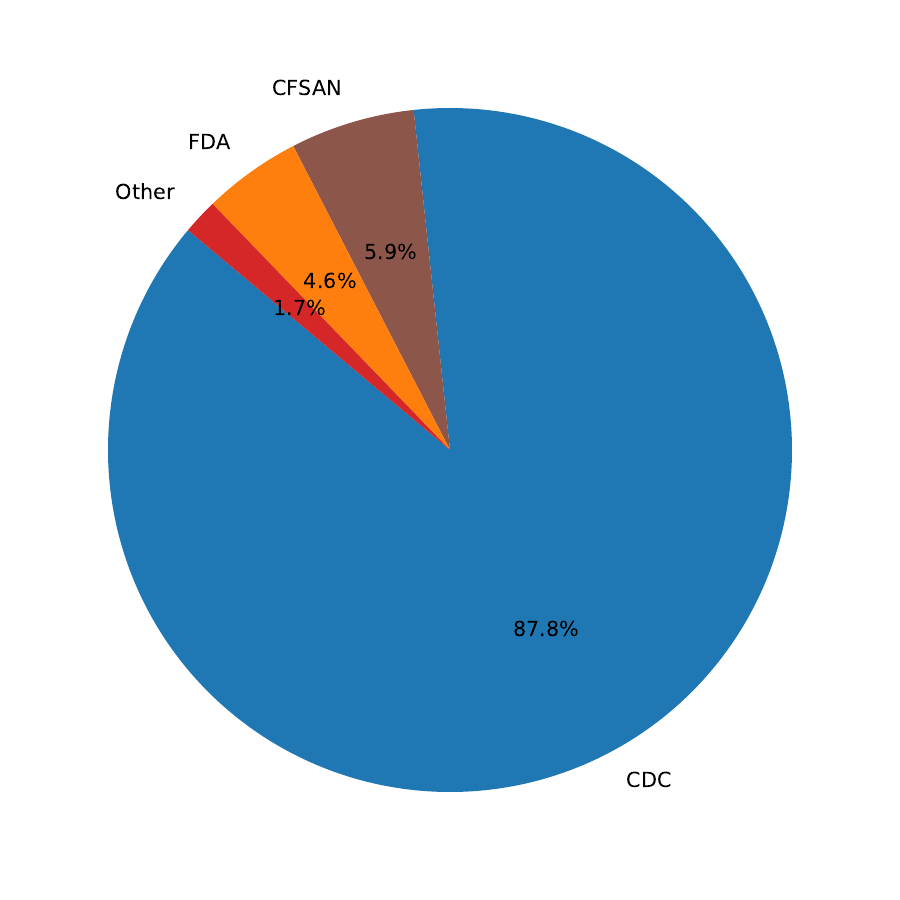}
\end{minipage}
\caption{(NCBI) Pie charts for data submissions for \textit{E. Coli} (left), and \textit{Salmonella} (Right). Although the majority of samples for both pathogens were submitted by CDC, there is a significant percentage of samples that were submitted by other entities; approximately 32\% for \textit{E. Coli}, and 12\% for \textit{Salmonella}.}
\label{Fig5}
\end{figure}

\begin{figure}[h!]
    \centering
    \includegraphics[width=1\textwidth]{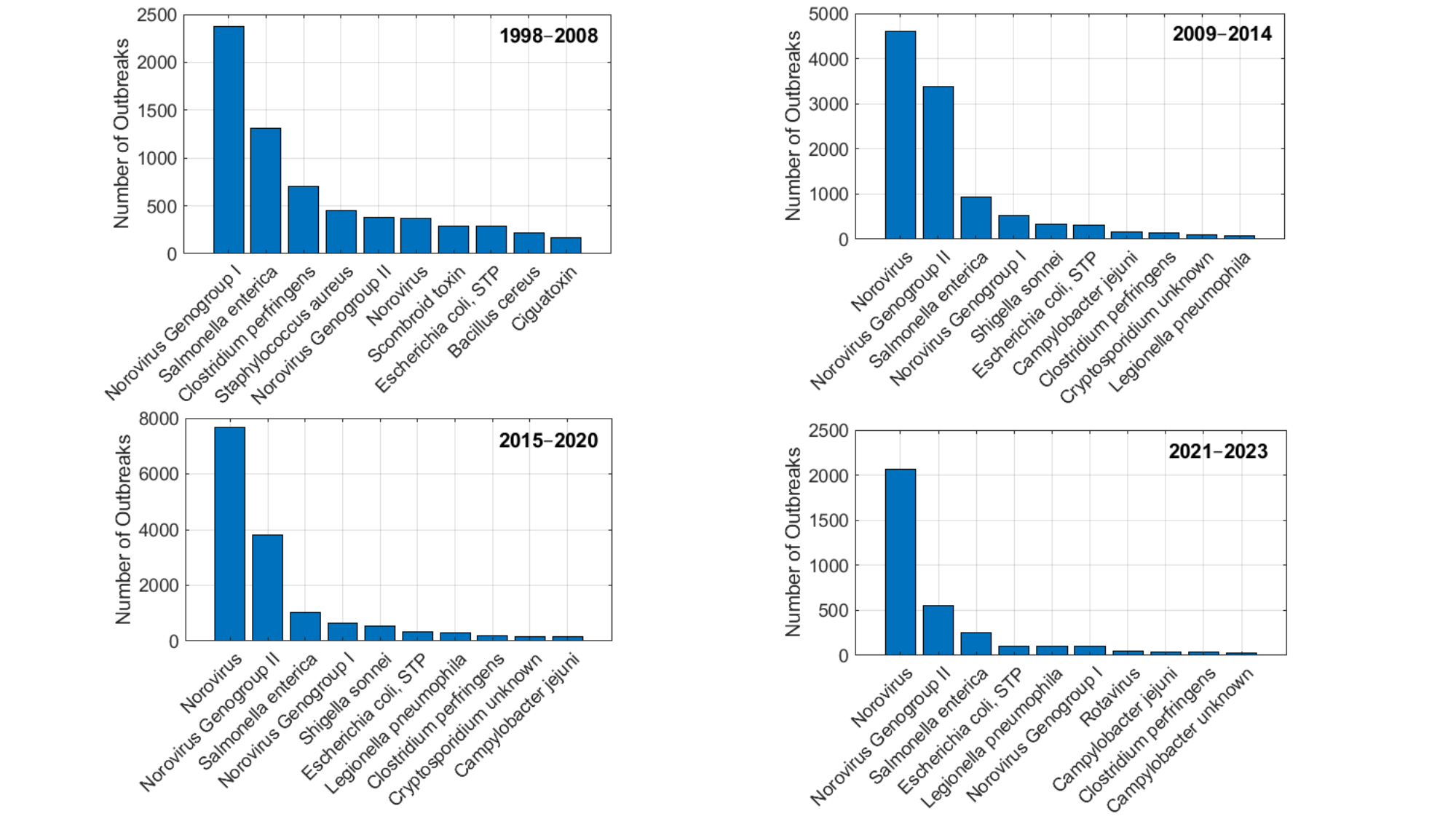}
    \caption{(NORS) Bar charts of more significant pathogens causing outbreaks for time intervals of 1998-2008, 2009-2014, 2015-2020, 2020-2023 obtained from NORS data}
    \label{fig:sample}
\end{figure}

\section{Temporal analysis of data}

\noindent
This section provides additional details, results, and diagnostics for the temporal analyses (e.g., model fit over time, change-point checks, forecast error profiles).

\begin{figure}[h!]
    \centering
    \includegraphics[width=1.0\textwidth]{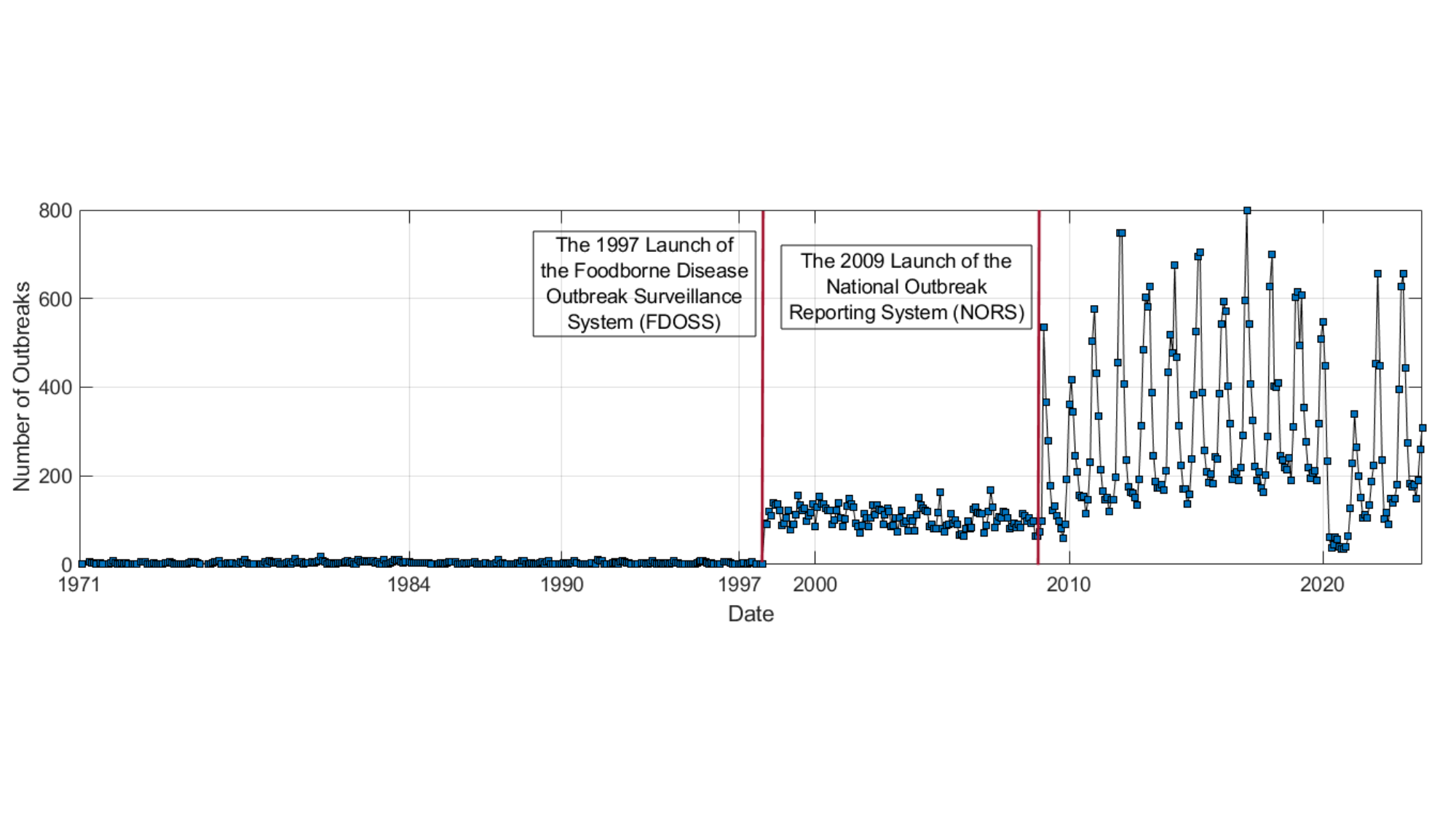}
    \caption{(NORS) In the time series data, there are two major jumps. The first jump is associated with the launch of the Foodborne Disease Outbreak Surveillance System in 1997, and the second is related to the launch of the National Outbreak Reporting System (NORS) in 2009. The jumps represent sharp increases in reporting efficiency for outbreaks, and do not necessarily represent a sharp increase in outbreaks occurring.
}
    \label{fig:sample}
\end{figure}

\begin{figure}[h!]
    \centering
    \includegraphics[width=1\textwidth]{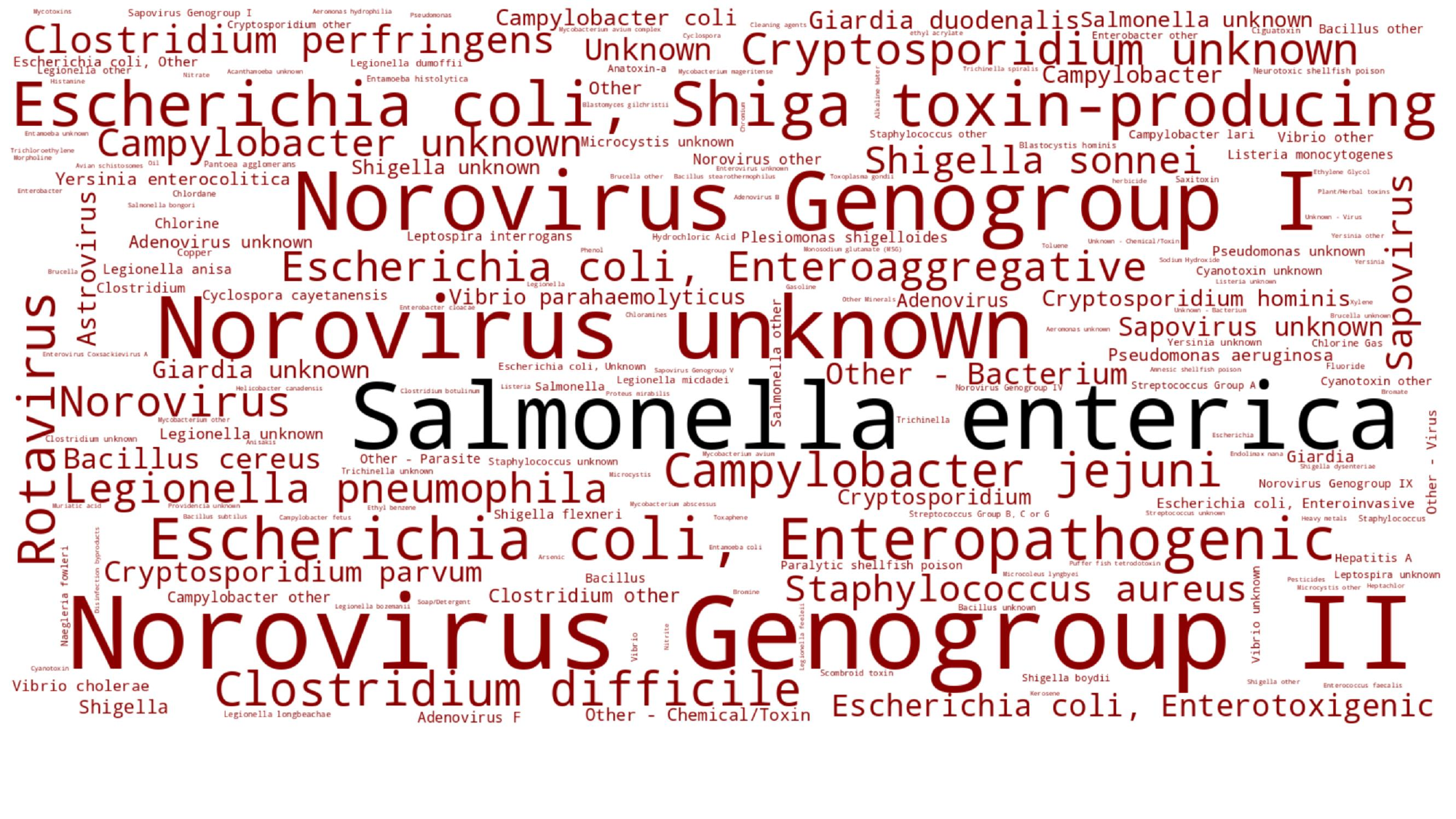}
    \caption{(NORS) Word Cloud of Etiologies. The word cloud shows that 'Norovirus unknown' and Norovirus Genogroup 1 and 2 are the main pathogens for outbreaks in the US. Thereafter, salmonella and E coli are the primary causes of outbreaks reported in the US.}
    \label{fig:sample}
\end{figure}

\begin{figure}[h!]
    \centering
    \includegraphics[width=1\textwidth]{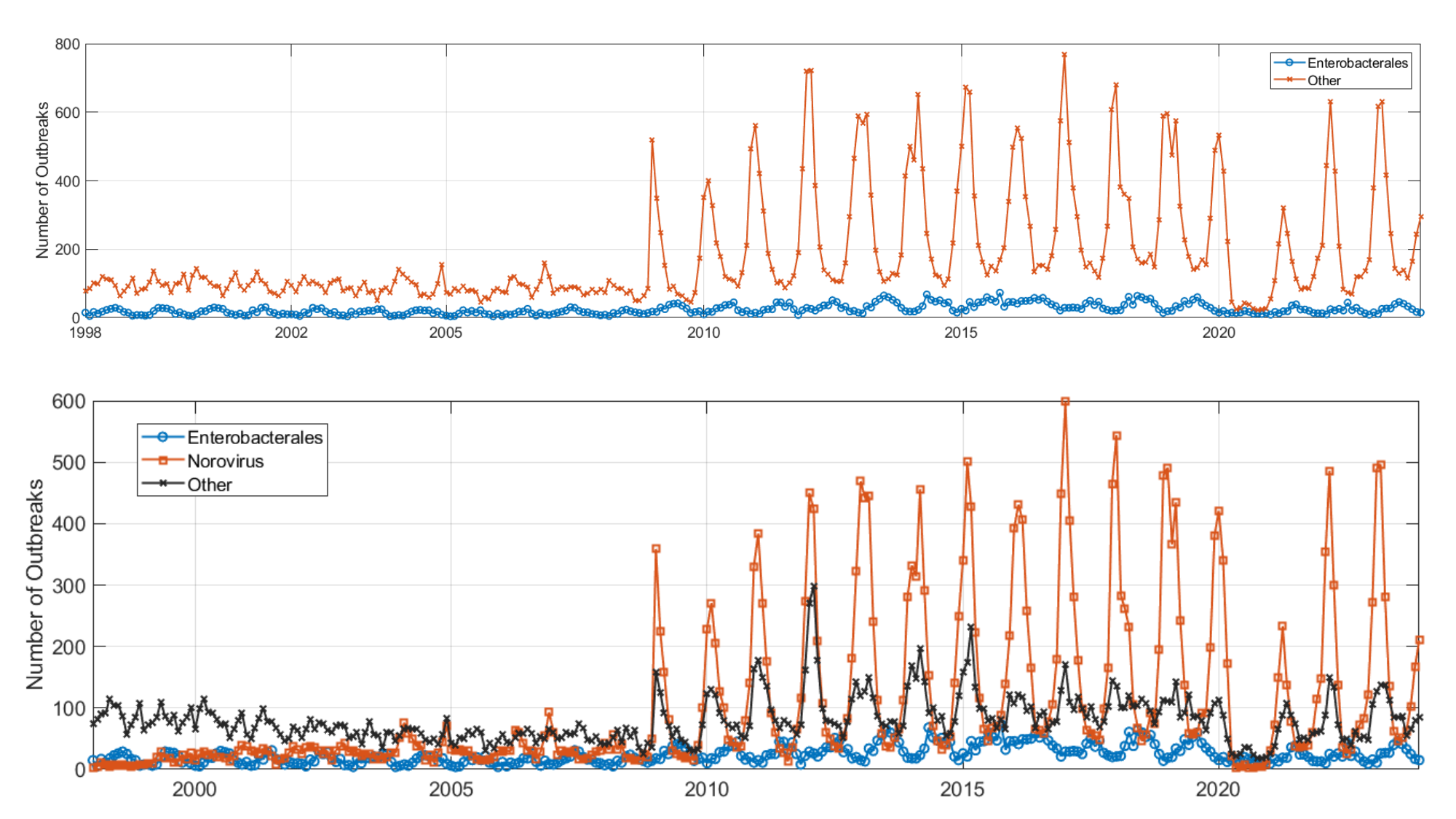}
    \caption{(NORS) Given that \textit{Salmonella}, \textit{E. Coli} are both members of the order \textit{Enterobacterales}, a comparison between time series of \textit{Enterobacterales} and \textit{Norovirus}-causing outbreaks in the US reveals interesting results. We can observe periodic behaviors in both time series The main difference is the peak values as  \textit{E. Coli}, \textit{Salmonella} mainly have peak values during the Summer, where whereas \textit{Norovirus} mostly occurs during Winter/Spring.}
    \label{fig:sample}
\end{figure}

\begin{figure}[htbp]
\includegraphics[width=0.5\linewidth]{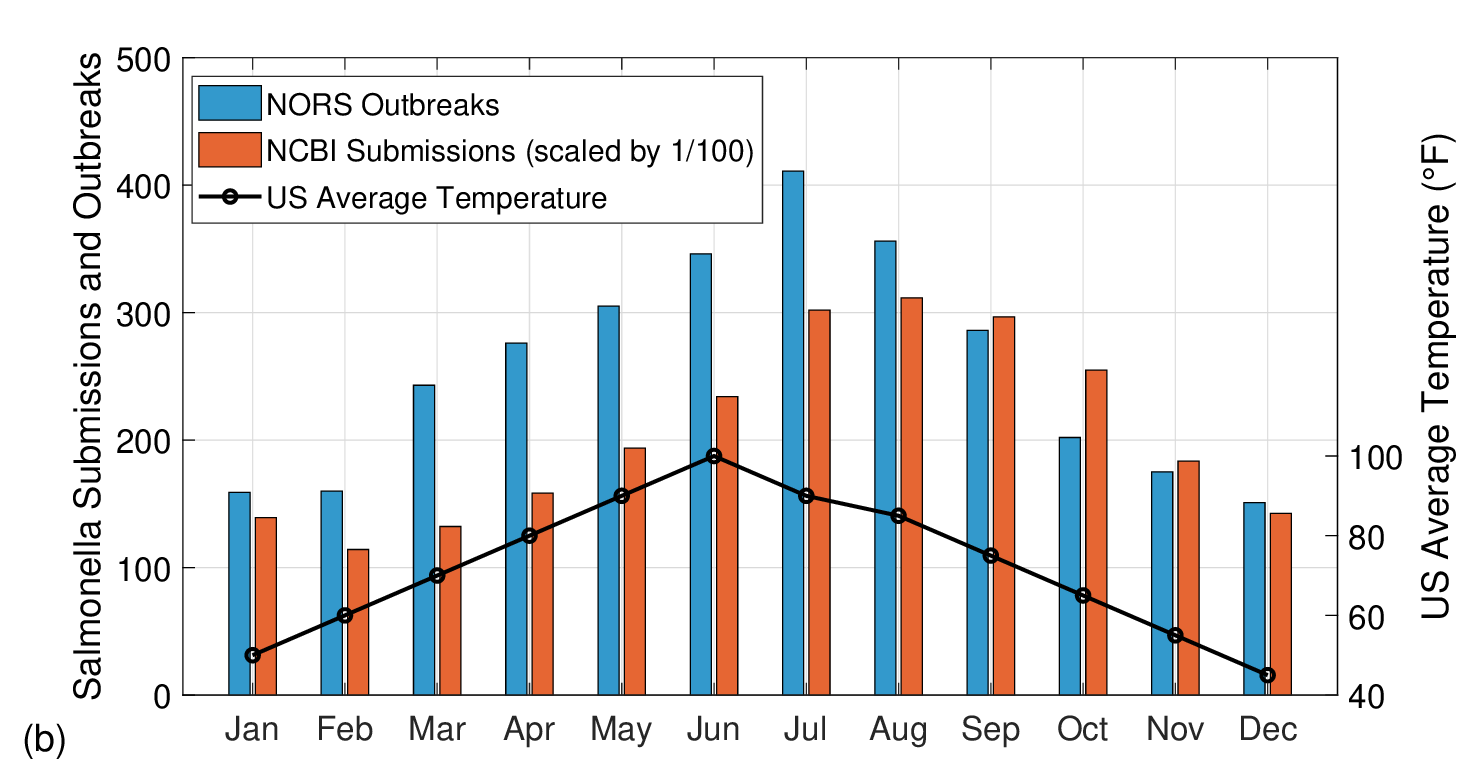}
        \includegraphics[width=0.5\linewidth]{S6.eps}
        \caption{(NCBI and NORS) Figures a) and b) provide a follow up to the claim of having higher numbers of outbreaks during Summer, where it can be seen that for both pathogens the numbers are much higher during June with a possible delay in reporting the outbreaks which are pushed to July}
    \label{fig:two-panels0}
\end{figure}

\begin{figure}[h!]
    \centering
    \includegraphics[width=1\textwidth]{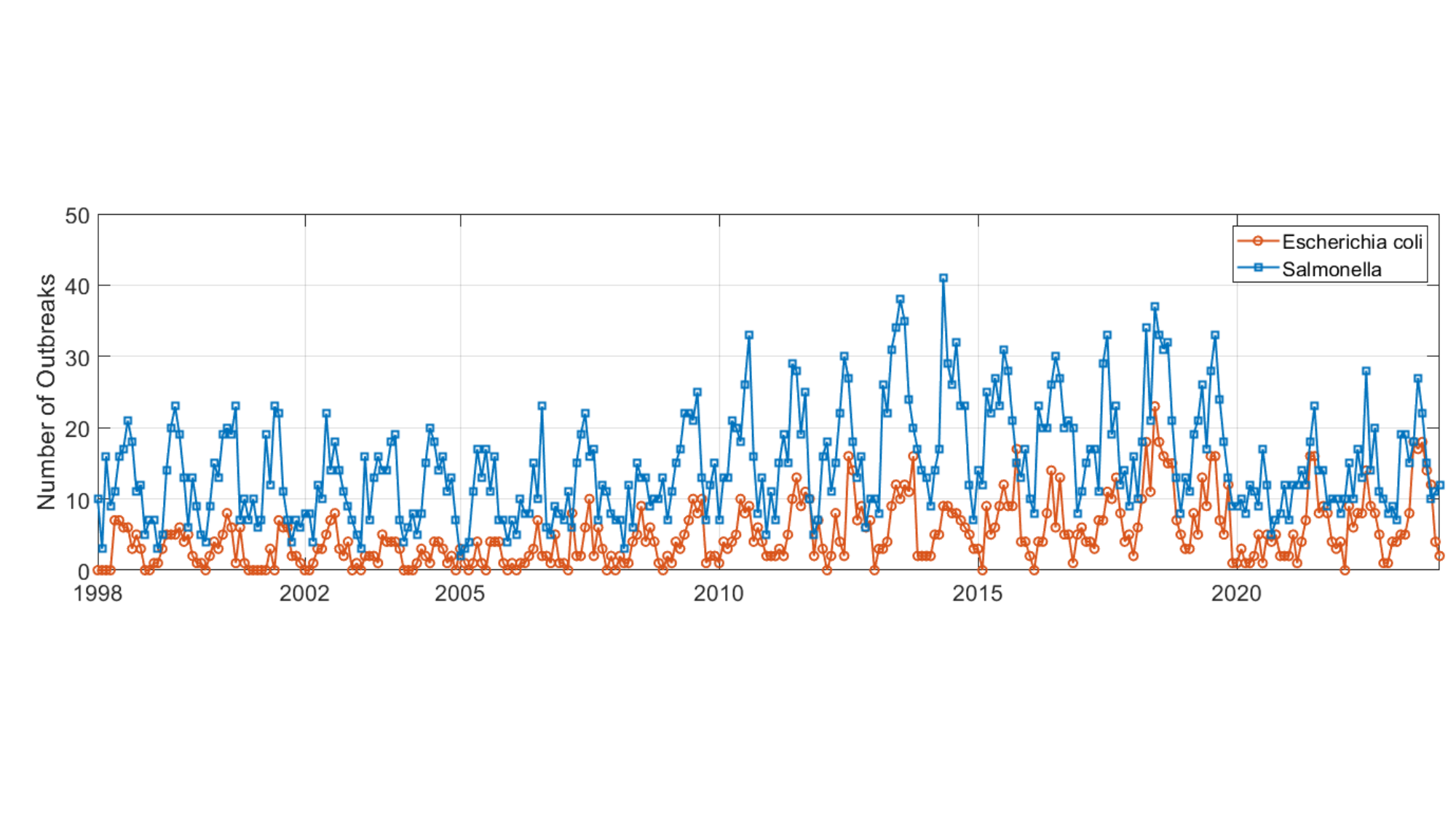}
    \caption{(NORS) This figure shows that the number of outbreaks caused by \textit{Salmonella} or two to three times higher than those caused by \textit{E. Coli}. We can still observe synchronized peaks between outbreaks caused by these two pathogens}
    \label{fig:sample}
\end{figure}

\section{Spatial analysis of data}

\noindent
This section contains supplementary maps and spatial diagnostics.
\begin{figure}[h!]
    \centering
    \includegraphics[width=1\textwidth]{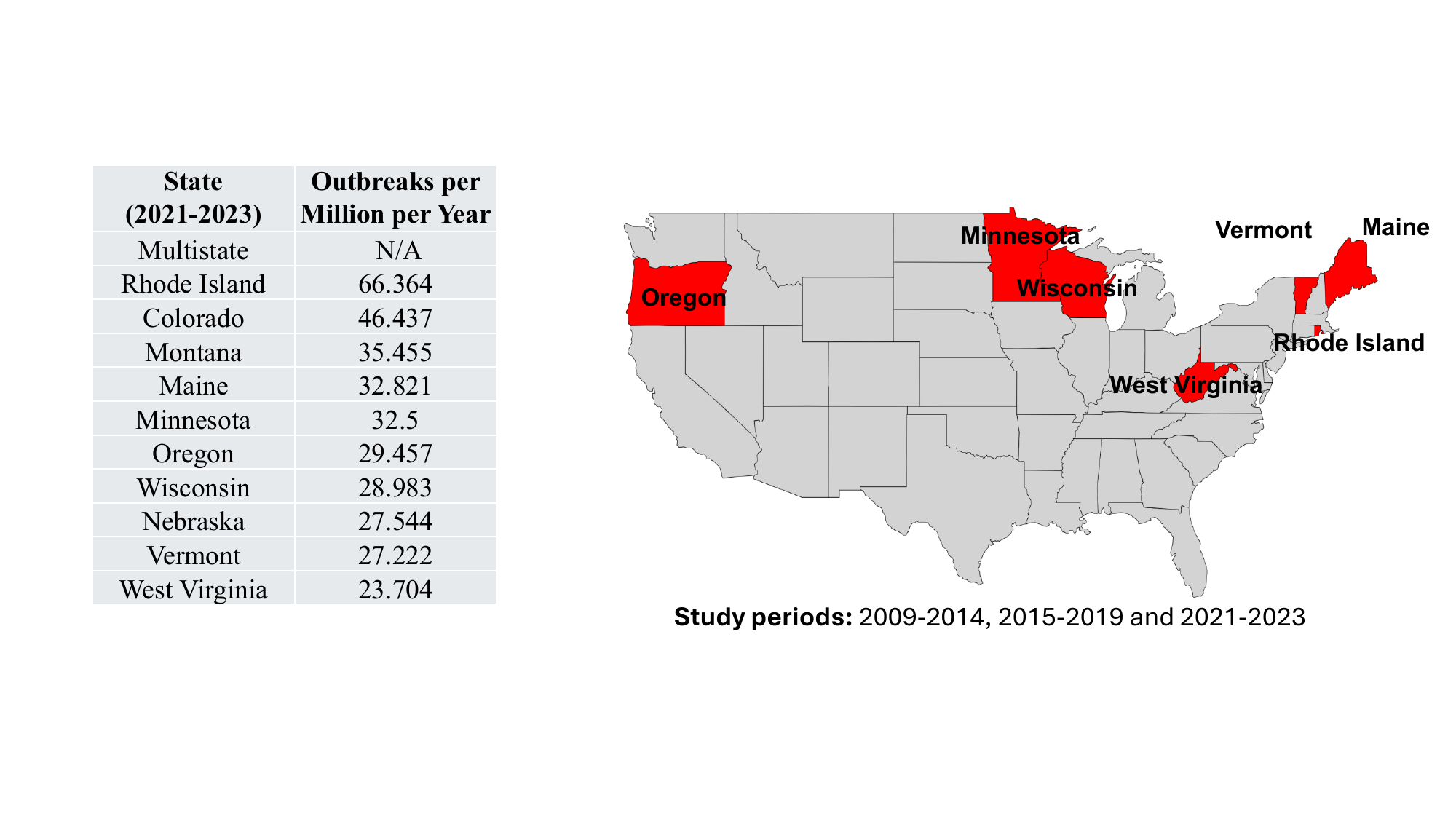}
    \caption{(NORS) The states with the highest number of reported outbreaks during 2009 to 2023 are displayed. Focusing on the last two years of the study, it can be seen that by scaling the number of outbreaks according to the population of each state V identify states with per capita highest outbreaks which has been represented in the table.
}
    \label{fig:sample}
\end{figure}

\begin{figure}[h!]
    \centering
    \includegraphics[width=1\textwidth]{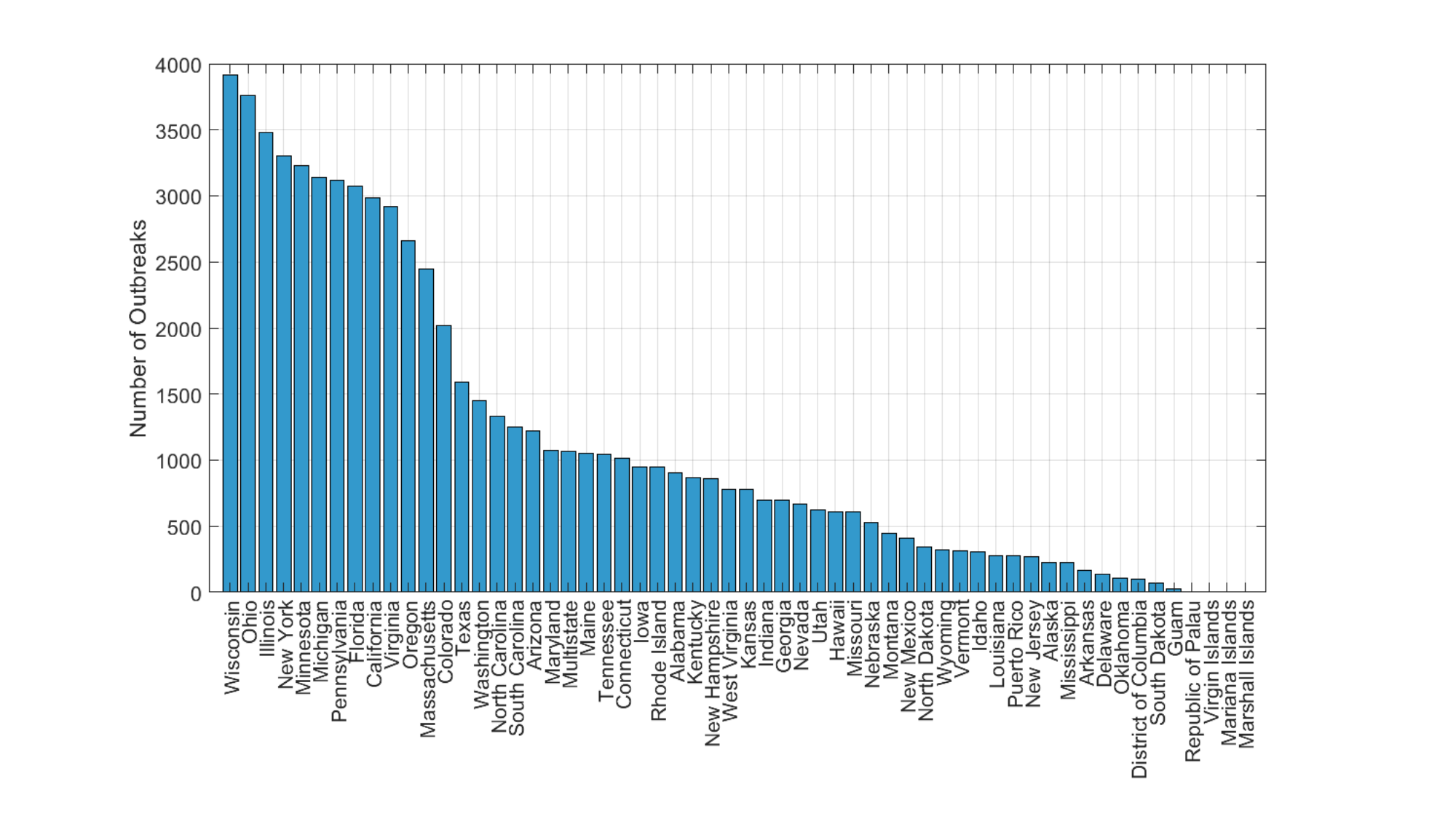}
    \caption{(NORS) The states with the highest number of reported outbreaks during 2009 to 2023.}
    \label{fig:sample}
\end{figure}

\begin{figure}[h!]
    \centering
    \includegraphics[width=1\textwidth]{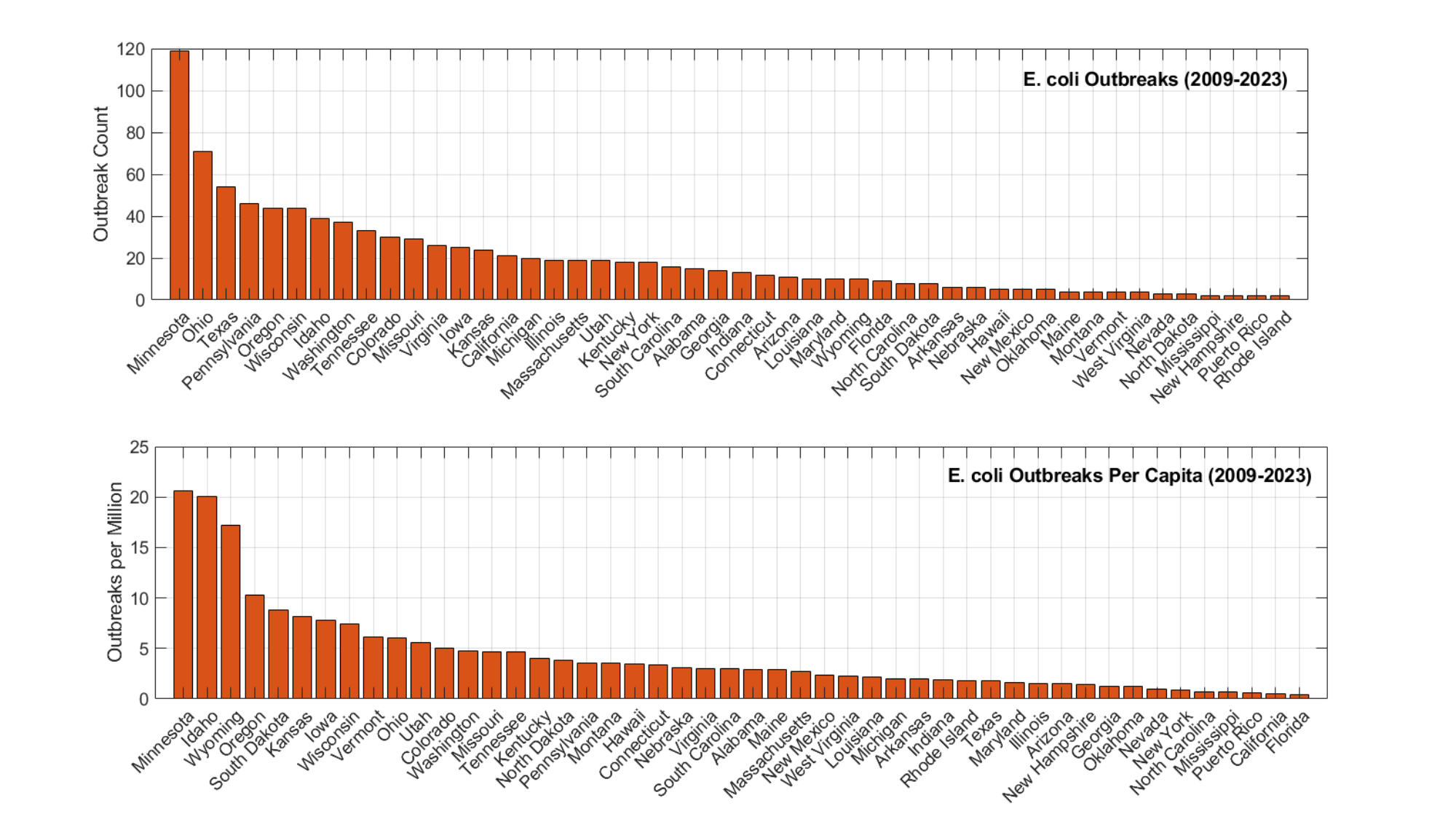}
    \caption{(NORS) The states with the highest number of reported outbreaks associated with \textit{E. Coli} from 2009 to 2023. Note that the top plot and each figure corresponds to reported numbers of outbreaks and the bottom plot in each figure corresponds to numbers of outbreaks per capita.}
    \label{fig:S9}
\end{figure}
\begin{figure}[h!]
    \centering
    \includegraphics[width=1\textwidth]{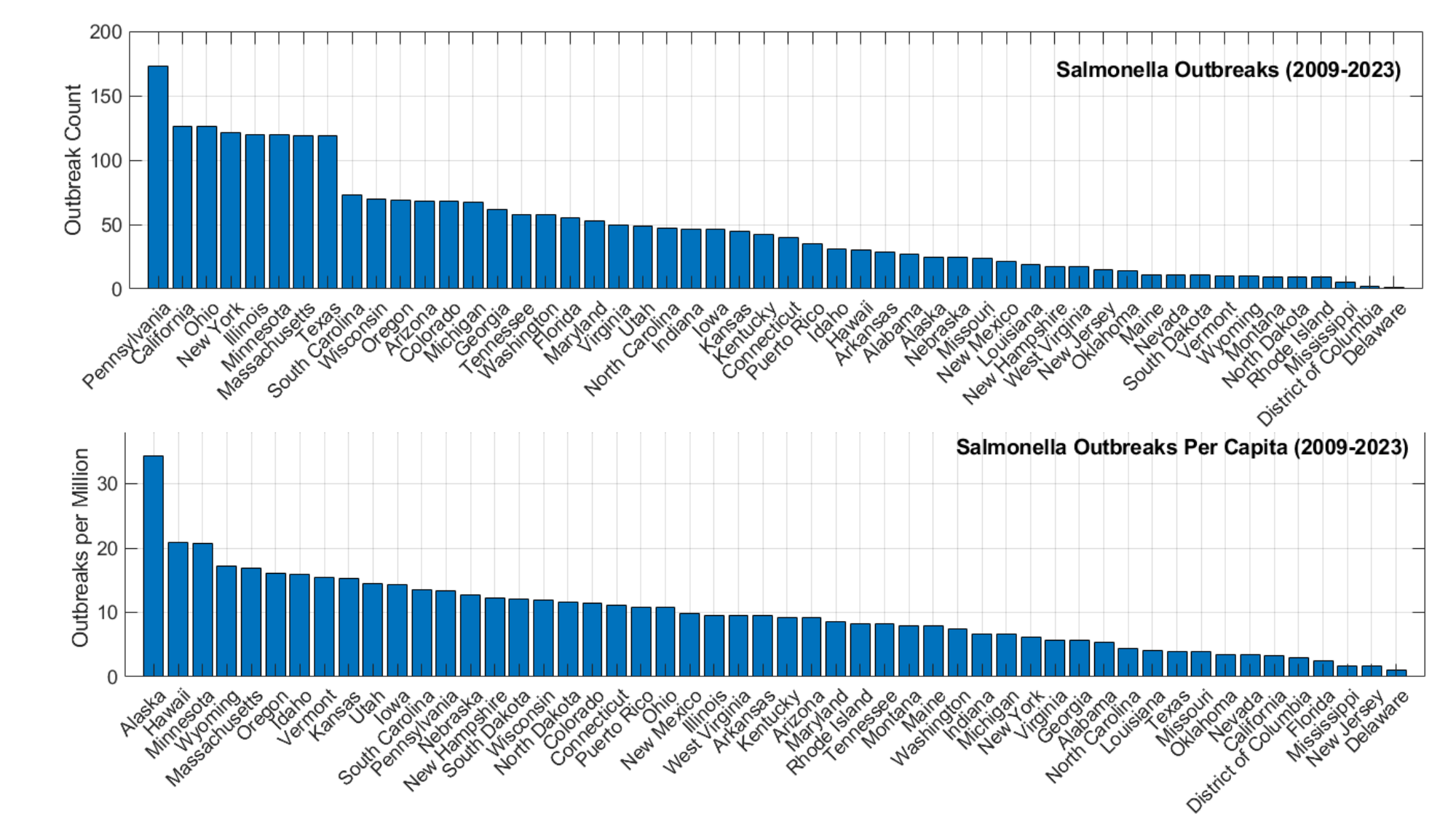}
    \caption{(NORS) The states with the highest number of reported outbreaks associated with \textit{Salmonella} from 2009 to 2023. Note that the top plot and each figure corresponds to reported numbers of outbreaks and the bottom plot in each figure corresponds to numbers of outbreaks per capita of the reported state.}
    \label{fig:S10}
\end{figure}

\begin{figure}[h!]
    \centering
    \includegraphics[width=1\textwidth]{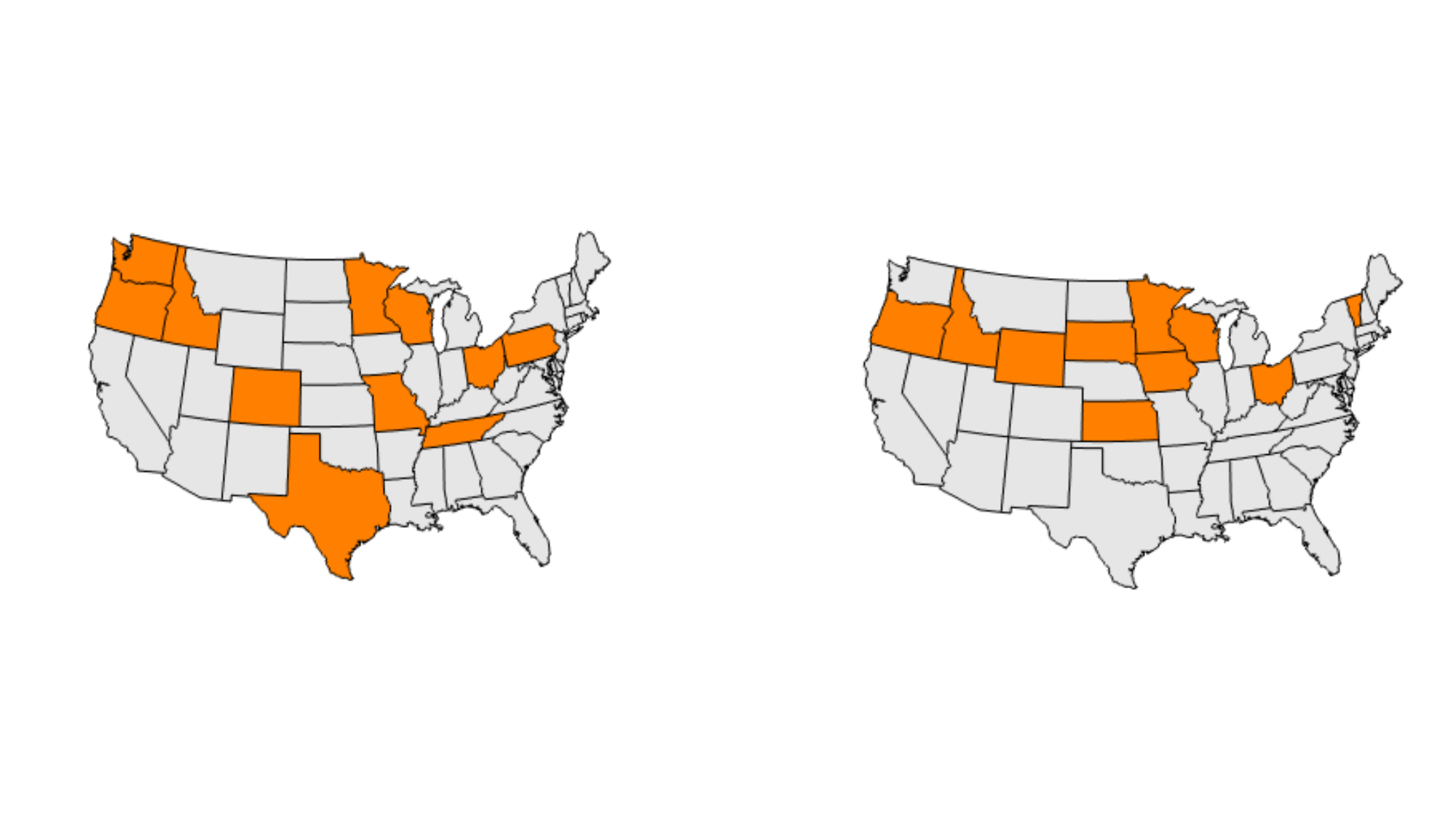}
    \caption{(NORS) US map highlighting the top states for \textit{E. Coli} based on Figure \ref{fig:S9}}
    \label{fig:sample}
\end{figure}
\begin{figure}[h!]
    \centering
    \includegraphics[width=1\textwidth]{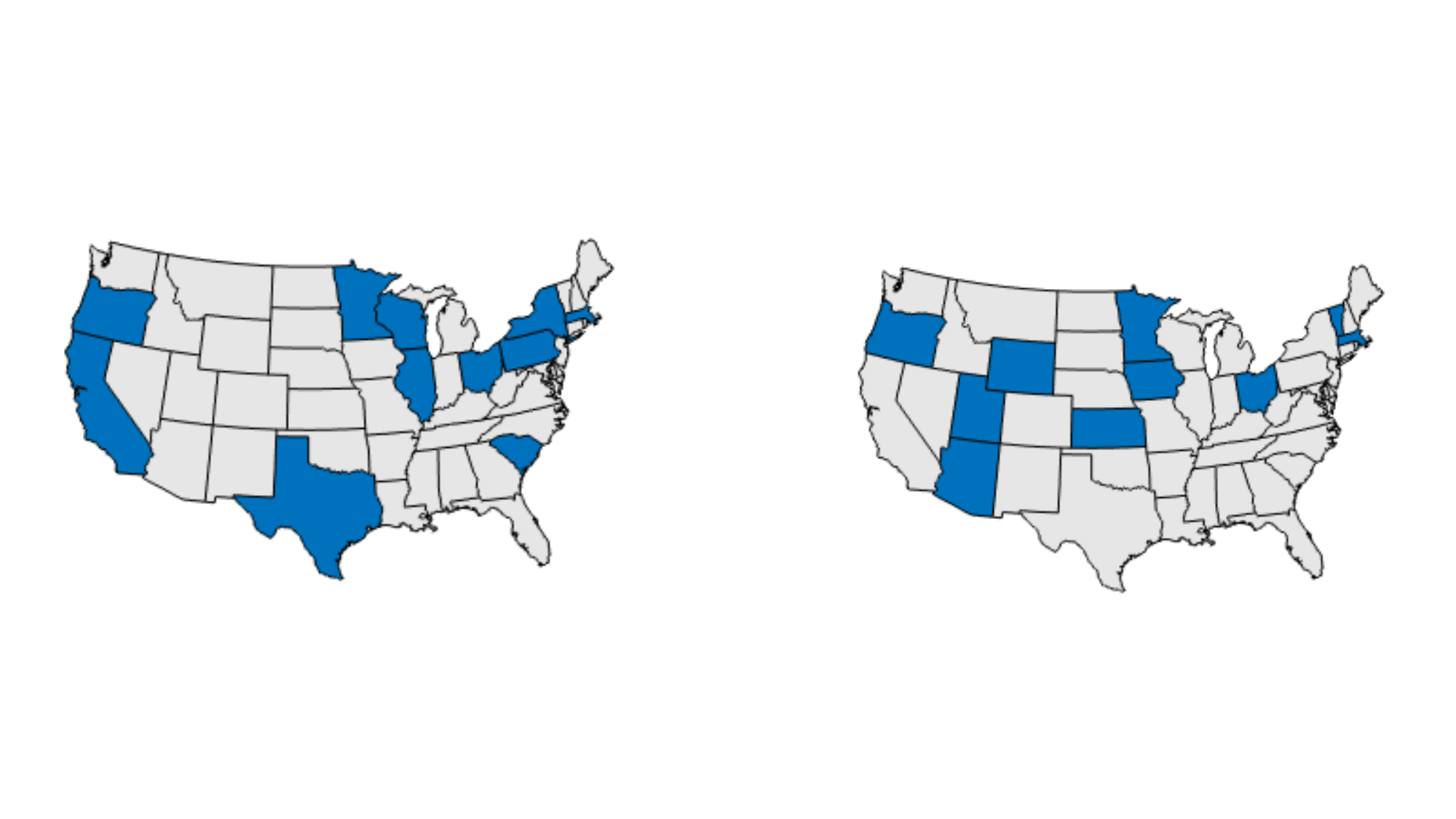}
    \caption{(NORS) US map highlighting the top states for \textit{Salmonella} based on Figure \ref{fig:S10}}
    \label{fig:sample}
\end{figure}